\documentclass[aps,prl,reprint,superscriptaddress]{revtex4-1}

\usepackage{graphicx}
\usepackage{import}
\usepackage{amsmath}
\usepackage{amssymb}
\usepackage{float}
\usepackage{bigints}
\usepackage{physics}
\usepackage[normalem]{ulem}
\usepackage[colorlinks=true,linkcolor=blue,citecolor=blue,urlcolor=blue]{hyperref}
\usepackage[capitalise]{cleveref}
\usepackage{mwe}
\usepackage[dvipsnames]{xcolor}
\usepackage[normalem]{ulem}

\usepackage{pdfpages}
\makeatletter
\AtBeginDocument{\let\LS@rot\undefined}
\makeatother

\begin{document}  


\title{Simulating multi-node Weyl semimetals in a Mixed Floquet lattice}

\author{Goutham Vinjamuri}
\affiliation{Department of Physics, Indian Institute of Science Education and Research (IISER) Tirupati, Tirupati 517619, India}
\affiliation{Department of Physics, Indian Institute of Science Education and Research Bhopal, Bhopal Bypass Road, Bhauri, Bhopal 462 066, Madhya Pradesh, India}

\author{Ashutosh Dubey}
\affiliation{Department of Physics, Indian Institute of Science Education and Research (IISER) Tirupati, Tirupati 517619, India}

\author{Ankur Das}
\email{ankur@labs.iisertirupati.ac.in}
\affiliation{Department of Physics, Indian Institute of Science Education and Research (IISER) Tirupati, Tirupati 517619, India}

\begin{abstract}
In this work, we simulate Weyl semimetal phases with multi-nodes containing $2$, $4$, $6$, and $8$ Weyl nodes, in a mixed Floquet synthetic lattice generated by driving a one-dimensional chain with two incommensurate drives. In contrast to the integer-quantized response of Chern insulators, we obtain a momentum-resolved power-transfer signature, where coupling the drives to fixed momenta results in quantized power-transfer for each 2D slice corresponding to that momentum, thereby locating the Weyl nodes in the low-frequency limit. Further, we explicitly show for the Weyl phase with 2 nodes, in the adiabatic limit, the half-filled state exchanges energy between the two drives at a rate that converges to the projected Weyl-node separation equal to the Fermi-arc length of the Weyl semimetal.
\end{abstract}
\maketitle

\textit{Introduction---}
Topology in electronic band structures has become one of the central themes of modern condensed matter physics, leading to major advances in understanding topological insulators, semimetals, superconductors, and related phases \cite{RevModPhys.82.3045, RevModPhys.90.015001, RevModPhys.83.1057}. Yet many idealized topological Hamiltonians remain difficult to realize directly in solid-state materials. Periodic driving offers a complementary route by engineering effective Hamiltonians that realize topological insulators, topological superconductors, Floquet-Weyl semimetals, and phases without static analogues~\cite{cayssol2013floquet, rudner2020band, oka2019floquet, PhysRevResearch.4.033213, hubener2017creating}. In such driven systems, topology controls nonequilibrium responses, allowing experimentally accessible observables to encode topological invariants~\cite{PhysRevLett.113.236803, PhysRevX.7.041008, PhysRevB.99.064306, PhysRevB.111.035431}.
This framework further enables the construction of synthetic dimensions, which provide a versatile route for engineering quantum systems beyond the constraints of ordinary spatial geometry~\cite{ozawa2019topological, arguello2024synthetic, yu2025comprehensive}. 
These systems can often be mapped onto effective hopping models in the corresponding lattice~\cite{PhysRevX.7.041008} (often called Floquet lattice), allowing concepts from lattice band theory and topological matter to be realized and probed dynamically. Synthetic-dimensional topological physics has therefore become a rapidly developing direction, with theoretical and experimental progress in ultracold atomic gases~\cite{PhysRevLett.108.133001, Mancini, Stuhl} and photonic platforms~\cite{Yuan:18, RevModPhys.85.299}

Floquet synthetic dimensions provide a particularly useful setting for simulating higher-dimensional topological systems~\cite{PhysRevB.106.144203, prg5-3srn, PhysRevLett.126.163602}. For example, a spin-$1/2$ system subjected to two incommensurate drives can effectively realize a half Bernevig--Hughes--Zhang (BHZ) model in the Floquet lattice~\cite{PhysRevX.7.041008}. The corresponding synthetic quasienergy bands are characterized by integer Chern numbers; zero for trivial bands and nonzero for topological bands~\cite{PhysRevB.99.064306}. Beyond static band topology, such systems exhibit dynamical signatures including half-integer quantized power conversion between the drives~\cite{PhysRevLett.125.100601}, with extensions to dissipative~\cite{PhysRevResearch.6.033124} and interacting settings~\cite{PhysRevResearch.2.022023}. These developments are complemented by experimental realizations in driven synthetic platforms~\cite{sridhar2024quantized, PhysRevLett.125.160505}.

\begin{figure}[t]
\centering
\includegraphics[width=\columnwidth]{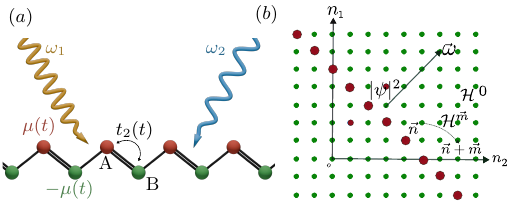}
\caption{(a) Schematic of a Rice--Mele chain coupled to two oscillating magnetic fields at frequencies $\omega_1$ and $\omega_2$. (b) Corresponding Floquet lattice: the green sites are labeled by $\vec{n}=(n_1,n_2)$, while $\vec{\omega}$ denotes the effective electric field. The brown-dot size encodes the wave-function weight: it decreases along $\vec{\omega}$, indicating localization, but remains nearly unchanged in the transverse direction, where the state is extended. The onsite term $\mathcal{H}^{0}$ and hopping term $\mathcal{H}^{\vec{m}}$ are also indicated.}
\label{fig:schematic}
\end{figure}

Recently, gapless topological phases, such as Weyl semimetals with two Weyl nodes, have been studied in the Floquet lattice~\cite{qi2021universal}. In particular, it shows that the energy pumping between two drives near the Weyl nodes has two contributions: a topological contribution arising from adiabatic dynamics and an excitation contribution arising from nonadiabatic quantum dynamics near the Weyl nodes. To obtain quantized energy pumping, the authors explicitly discussed mechanisms for suppressing the excitation contribution. Despite this progress, it remains unclear how Weyl semimetals with multiple Weyl nodes can be realized in this setup. It is also unexplored how the transition between different Weyl phases is encoded in the energy transfer between the two drives, and information about the Fermi-arc length can be extracted solely from the energy transfer between the drives. In this letter, we answer these questions by further studying more exotic Weyl phases with four, six, and eight nodes in a mixed Floquet lattice.  We find quantized topological frequency conversion only between the gapless points with opposite chiralities. More specifically, we show that reducing driving frequency leads to better matching of Fermi-arc length extracted from energy transfer between two drives and the one coming from the analytical formulae. Additionally, to extract information about the Fermi-arc length from the energy transfer between two drives, we explicitly show that a Rice–Mele-type chain coupled to two incommensurate drives exhibits non-quantized energy transfer in the low-frequency limit. However, the rate of energy transfer, i.e., the slope of the energy transferred by one drive with respect to time, directly provides information about the Fermi-arc length.

\textit{Floquet lattice formalism---} When a quantum system is subjected to multiple incommensurate driving frequencies, the time-dependent Hamiltonian becomes quasiperiodic in time. The conventional Floquet theorem for a single periodic drive can then be generalized by treating the phase associated with each drive as an independent variable. The Hamiltonian may be expanded as $\mathcal{H}(\vec{\theta}_t)=\sum_{\vec{n}}\mathcal{H}^{\vec{n}}e^{-i\vec{n}\cdot\vec{\theta}_t}$, where $\vec\theta_t = (\theta_{1t}, \theta_{2t})$ and $\vec n = (n_1, n_2)$. Here $\theta_{i t} = \omega_i t + \theta_{0i}$ for $i\in1,2$. 
$\theta_{it}$, $\theta_{0i}$, and $\omega_i$ denote, respectively, the instantaneous phase, initial phase, and frequency of the $i$-th drive, while the integer $n_i$ labels the number of photons absorbed from or emitted into that drive. We choose $\omega_1/\omega_2=\beta=(1+\sqrt{5})/2$, the golden ratio.\\
The solution of the Schr\"odinger equation for a multifrequency-driven system can be written as $\ket{\psi(\vec{\theta}_{t})}=e^{-i\varepsilon(\vec{\theta}_{0})t}\sum_{\vec{n}}e^{-i\vec{n}\cdot\vec{\omega}t}\ket{\phi^{\vec{n}}}$ \cite{PhysRevX.7.041008}, where $\varepsilon(\vec{\theta}_{0})$ is the quasienergy of the \textit{Floquet synthetic band}. 
The initial phases \((\theta_{01},\theta_{02})\in(0,2\pi]\) define the Brillouin zone of this synthetic band. 
Substituting this ansatz into the Schr\"odinger equation gives
\begin{align}
   \sum_{\vec{m}}\left[ \mathcal{H}^{\vec{n}-\vec{m}}-\vec{n}\cdot\vec{\omega}\delta_{\vec{n},\vec{m}}\right]\ket{\phi^{\vec{m}}}
   =
   \varepsilon(\vec{\theta}_{0})\ket{\phi^{\vec{n}}},
   \label{eq:1}
\end{align}
which can be interpreted, in analogy with the Wannier--Stark ladder, as a tight-binding problem with an electric field in the \textit{Floquet lattice}. Here \(\vec{n}=(n_1,n_2)\) labels the lattice position generated by the two drives, \(\mathcal{H}^{\vec{n}}\) denotes hopping amplitude in this synthetic lattice, and \(\mathcal{H}^{\vec{0}}\) is the onsite term. The second term, \(\vec{n}\cdot\vec{\omega}\), acts as a linear potential generated by an effective electric field \(\vec{\omega}\), as illustrated in \cref{fig:schematic}(b). The dimension of the Floquet lattice is therefore equal to the number of independent incommensurate drives. A one-dimensional chain of two-level systems driven at two incommensurate frequencies thus realizes a mixed Floquet lattice with one real and two synthetic dimensions.

\textit{Model---}
Following the prescription to construct a synthetic Floquet lattice as described previously \cite{PhysRevX.7.041008}, we construct our mixed synthetic Floquet lattice by replacing two of the three momentum variables (namely $k_y$ and $k_z$) with drive variables.
In particular, we start from a time-reversal-broken Weyl-semimetal Hamiltonian~\cite{abdulla2022}, leading to the driven Hamiltonian
\begin{equation}
\begin{aligned}
\mathcal{H}/\eta &= \vec{\sigma}\cdot\vec{B}(t),\\
\vec{B}(t) &= 
\begin{pmatrix}
2\left[M(k_x)-t_y\cos(\omega_1 t+\theta_{01})-t_z\cos(\omega_2 t+\theta_{02})\right]\\
2t_y\sin(\omega_1 t+\theta_{01})\\
2t_z\sin(\omega_2 t+\theta_{02})
\end{pmatrix},
\end{aligned}
\label{eq:2}
\end{equation}
where $M(k_x)=m-t_x\cos k_x$ and $\eta$ is a scaling parameter. The corresponding static Bloch Hamiltonian can be recovered by identifying $\omega_1 t+\theta_{01}\to k_y$ and $\omega_2 t+\theta_{02}\to k_z$. The resulting phase diagram depends on $t_x/t_y$ and $m/t_y$ \cite{abdulla2022}. Throughout this work, we set $t_y=t_z=1$ for the two-node Weyl phase, unless stated otherwise.\\

\begin{figure*}
  \centering
  \includegraphics[width=0.95\textwidth]{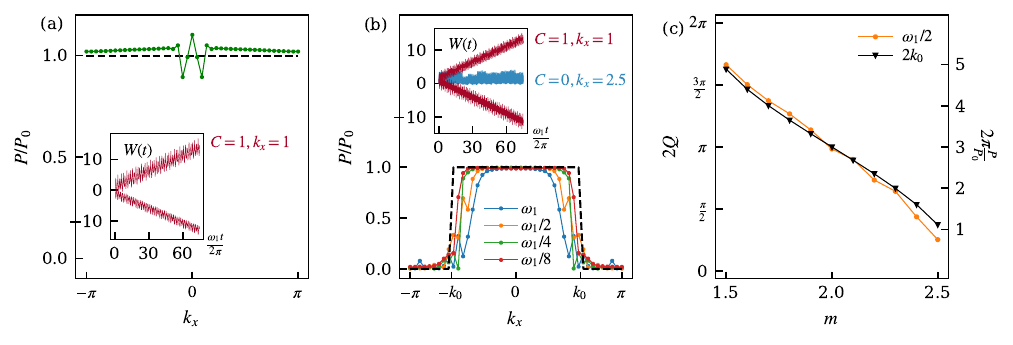}
  \caption{Momentum-resolved and real-space signatures of Weyl topology in the $\mathrm W_2$ phase. Panels (a) and (b) compare the Chern number of each 2D slice extracted from the long-time-averaged inter-drive power transfer $P/P_0$ (coloured solid lines with circles), with $C(k_x)$ calculated for $\mathcal{H}(k_x,\theta_1,\theta_2)$ over the $(\theta_1,\theta_2)$ torus using the Fukui--Hatsugai--Suzuki method~\cite{fukui2005chern} (black dotted line). (a) Results for $m=0.8$, $t_x=0.6$, and $\eta=5$ in the LCI phase; the inset shows the work done by the two drives at $k_x=1$. (b) Results for $m=1.8$, $t_x=0.6$, and $\eta=1$ in the $\mathrm{W}_2$ phase. The colored curves correspond to successively lower driving frequencies and their convergence toward $C(k_x)$ in the zero-frequency limit; the inset shows the work done by the two drives at $k_x=1$ (red) and $k_x=2.5$ (blue). (c) The total real-space response, calculated using the Fourier-transformed Hamiltonian in \cref{eq:7} for $t_x=0.7$, as a function of $m$ in the $\mathrm{W}_2$ phase. For frequency $\omega_1/2$, the quantity $2\pi \frac{P}{P_0}$ (orange circles)  shows very good agreement with the theoretically calculated Fermi-arc length $2Q$ (black triangles) of the Weyl semimetal.}
  \label{fig:CvsK}
\end{figure*}

Furthermore, the Fourier components generated by the $i$-th drive, $e^{\pm i\omega_i t}$, naturally act as nearest-neighbor hopping operators along the corresponding Floquet-lattice direction. Specifically, they connect photon-number sectors $n_i$ and $n_i\mp 1$. Thus, the multifrequency Schr\"odinger equation can be viewed as a tight-binding problem on the Floquet lattice. For our specific model, the Schr\"odinger equation in the mixed Floquet lattice is given by
\begin{equation}
\begin{aligned}
  i\partial_{t}\psi_{n_{1}, n_{2}}(k_{x}) &=  2\left(m -t_{x}\cos(k_{x}) - n_{1}\omega_{1} - n_{2}\omega_{2}\right)\psi_{n_{1},n_{2}}(k_{x}) \\ 
  &+ (-i t_{y} \sigma_{y} - t_{y}\sigma_{x})e^{i\theta_{01}}\psi_{n_{1} -1, n_{2}}(k_{x})\\
  &+ (i t_{y} \sigma_{y} - t_{y}\sigma_{x})e^{-i\theta_{01}}\psi_{n_{1} + 1, n_{2}}(k_{x})\\
  &+ (-i t_{z} \sigma_{z} - t_{z}\sigma_{x})e^{i\theta_{02}}\psi_{n_{1},n_{2} - 1}(k_{x})\\
  &+ (i t_{z} \sigma_{z} - t_{z}\sigma_{x})e^{-i\theta_{02}}\psi_{n_{1},n_{2} + 1}(k_{x}),
 \label{eq:3}
\end{aligned}
\end{equation}
where \(\psi_{n_1,n_2}(k_x)\) denotes the wave-function amplitude at the Floquet site \(\vec{n}=(n_1,n_2)\) for momentum \(k_x\). From the above equation, the Hamiltonian can be separated into two parts: a hopping term and a potential term arising from the effective electric field \(\vec{\omega}=(\omega_1,\omega_2)\). The hopping part of the Hamiltonian is translationally invariant in the Floquet lattice and can therefore be Fourier transformed. As a result, the total Hamiltonian takes the form \(\mathcal{H}=\sum_{\vec{q}} \mathcal{H}_{\vec{q}} n_{\vec{q}} + \sum_{\vec{n}} \mathcal{H}_{\vec{n}} n_{\vec{n}}\), where
\begin{equation}
\begin{aligned}
 \mathcal{H}_{\vec{q}} &= \left[2(m - t_{x}\cos(k_{x}) - t_{z}\cos(q_{1} + \theta_{01}) - t_{y}\cos(q_{2} + \theta_{02}) \right],\\
  \mathcal{H}_{\vec{n}} &= \vec{n}\cdot\vec{\omega},
   \label{eq:4}
\end{aligned}
\end{equation}
where \( n_{\vec{q}} \) and \( n_{\vec{n}} \) represent the occupation of the momentum site \(\vec{q}=(q_{1},q_{2})\) and the Floquet site \((n_{1},n_{2})\), respectively. In the above equation, \(\mathcal{H}_{\vec{q}}\) is the momentum-space representation of the Hamiltonian in the Floquet lattice. At \(t=0\), the momentum-space Hamiltonian generates the Floquet synthetic band structure for all values of \(\{\theta_{01},\theta_{02}\}\in(0,2\pi]\). The electric field \(\vec{\omega}\) generates the potential term \(\mathcal{H}_{\vec{n}}\), which evolves the momentum as \(\vec{q}=\vec{\omega}t\), and therefore leads to the evolution of the synthetic band structure.\\

\textit{Dynamical observable---}
{In the adiabatic regime, where the driving frequencies are smaller than the relevant synthetic-band gap, Landau-Zener transitions are suppressed, and the dynamics can be described semiclassically. For a synthetic band with dispersion \(\epsilon_{\vec q}\) and Berry curvature \(\Omega_{\vec q}\), the equations of motion are
\begin{equation}
   \dot{\vec{n}}=\nabla_{\vec q}\epsilon_{\vec q}+\vec{\omega}\times\Omega_{\vec q},
   \qquad
   \dot{\vec q}=\vec{\omega},
   \label{eq:5}
\end{equation}
where \(\vec{\omega}\) plays the role of an effective electric field in the Floquet lattice. The anomalous velocity \(\vec{\omega}\times\Omega_{\vec q}\) produces motion transverse to this field and is the temporal analog of chiral motion in quantum Hall systems. Since motion in the Floquet lattice changes the photon numbers \(\vec n=(n_1,n_2)\), it directly corresponds to energy exchange with the drives. For the \(i\)-th drive, \(dE_i/dt=\omega_i\,dn_i/dt\). The topological contribution from the anomalous velocity gives \(d\varepsilon_1/dt=-d\varepsilon_2/dt=\omega_1\omega_2\Omega_{\vec q}\), showing that Berry curvature mediates energy transfer between the two drives. When \(\vec q\) explores the full synthetic Brillouin zone, the long-time average becomes quantized as \( d\bar{\varepsilon}_1/dt=-d\bar{\varepsilon}_2/dt=\omega_1\omega_2 C/(2\pi)\), where \(C=(1/2\pi)\int d^2q\,\Omega_{\vec q}\) is the Chern number of the synthetic band.

The same energy transfer can be computed directly from the work operator associated with each drive~\cite{PhysRevX.7.041008},
\begin{equation}
    \frac{d\hat{W}_{i}}{dt}
    =
    \mathcal{U}^{\dagger}(t)
    \frac{d h_i(t)}{dt}
    \mathcal{U}(t),
    \label{eq:6}
\end{equation}
where \(\mathcal{U}(t)=\mathcal{T}\exp[-i\int_0^t\mathcal{H}(s)\,ds]\) is the time-evolution operator, \(\mathcal{H}(t)=\sum_i h_i(t)\cdot\sigma\), and \(i=1,2\) labels the two drives. The average work done by the \(i\)-th drive is obtained from \(\langle\psi(0)|\hat{W}_i|\psi(0)\rangle\), with \(|\psi(0)\rangle\) the initial state.}

\textit{Results and discussion---}
The objective of this work is to capture the topological properties of a three-dimensional time-reversal-symmetry-broken Weyl semimetal~\cite{abdulla2022} by driving a one-dimensional system with two incommensurate drives, as shown in \cref{fig:schematic}a, which can be equivalently described as a spin-half system in momentum space (\cref{eq:2}). 
\begin{figure*}
    \includegraphics[width=0.95\textwidth]{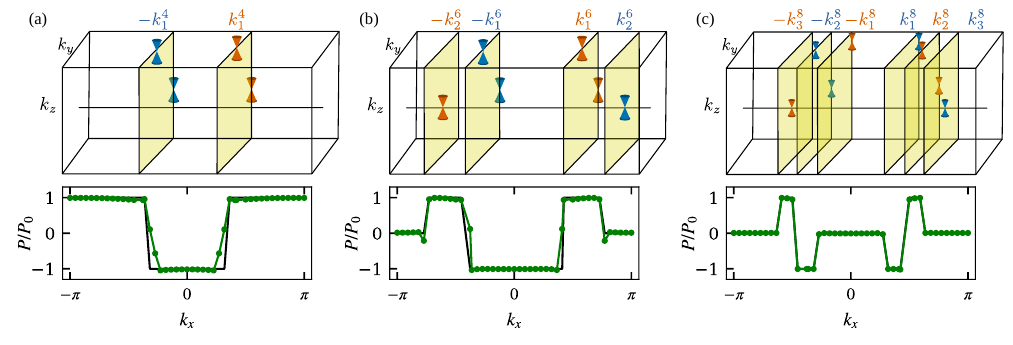}
    \caption{Comparing the expected Chern number as a function of $k_x$ with the net power transferred between the drives calculated in the $\mathrm W_4$, $\mathrm W_6$ and $\mathrm W_8$ phases. Each of the three plots are accompanied by a schematic representation the position of the Weyl nodes in the 3D Brillouin Zone. (a) In the $\mathrm W_4$ phase ($m=0.5$ and $t_x=1$), two Weyl nodes of the same chirality occour on one plane given by $k_x=-k^4_1$ and the other two Weyl nodes of the opposite chirality occour on the plane with $k_x=k^4_1$. The nodes are at $(\pm k^4_1, \pi, 0)$ and $(\pm k^4_1, 0, \pi)$. (b) In the $\mathrm W_6$ phase $(m=0.5, t_x=2)$, there are 6 Weyl nodes (3 pairs), which lie on 4 different planes as shown in the schematic. 
    The Weyl nodes are at $(\pm k^6_2, 0, 0)$, $(\pm k^6_1, \pi, 0)$ and $(\pm k^6_1, 0, \pi)$. (c) The $\mathrm W_8$ $(m=0.5, t_x=4)$ phases hosts 8 Weyl nodes (4 pairs) at locations $(\pm k^8_3, 0, 0)$, $(\pm k^8_2, \pi, 0)$, $(\pm k^8_2, 0, \pi)$ and $(\pm k^8_1, \pi, \pi)$. The locations of Weyl nodes in all of the Weyl phases discussed above are given by $\cos(k_x)=(m-(-1)^\mu t_y -(-1)^\nu t_z)/t_x$, where $\mu,\nu=0,1$ depending on whether $k_y$ and $k_z$ respectively take the values $0$ or $\pi$.}
    \label{fig:fig3}
\end{figure*}
We drive the directions perpendicular to the Weyl axis, so that the degeneracies along the momentum direction remain at $\cos(k_{0}) = (m - (-1)^{\mu}t_{y} - (-1)^{\nu}t_{z})/t_{x}$. Here $\mu,\nu=0,1$, depending on whether $k_y$ and $k_z$ are $0$ or $\pi$ \cite{abdulla2022}. The static $(3+0)$-dimensional Weyl semimetal exhibits Weyl phases with $2$, $4$, $6$, and $8$ nodes, along with a topological LCI (Layered Chern Insulator) phase and a trivial insulating phase, as shown in Ref.~\cite{abdulla2022}. {We vary the values of $m$ with $t_x/t_y=0.6$ to study either the LCI or the $\mathrm W_2$ phases. Similarly, we choose appropriate values of $t_x/t_y$ along the $m/t_y=0.5$ line in the phase diagram to study the $\mathrm W_4$, $\mathrm W_6$, and $\mathrm W_8$ phases. In any phase, we compute the power transfer between two drives, which is generically given by $P = n P_{0}$, where $P_0=\frac{\omega_1\omega_2}{\pi}$ and $n \in \mathbb R$. When $0.6<m/t_y<1.4$ along the line $t_x/t_y=0.6$, the static system is in the LCI phase. If the two-tone driven system (\cref{eq:2}) is initialized in this phase, the Floquet lattice also exhibits a non-trivial synthetic band structure. We compute the quantity $P(k_{x})/P_0$ and find that $n(k_{x}) = 1$ $\forall k_x \in [-\pi, \pi]$, which agrees with the Chern number calculated over the $(\theta_1, \theta_2)$ BZ for each of those $k_x$ momenta, using the Fukui--Hatsugai--Suzuki (FHS) method \cite{fukui2005chern} (black dotted lines in \cref{fig:CvsK}a). Moving right along the $t_x/t_y=0.6$ line, the system enters the $\mathrm W_2$ phase when $1.4<m/t_y<2.6$. Each point $(m,t_x)$ in this phase has two Weyl nodes along $k_x$, located at $k_x=\pm k_0$, where $k_0=\cos^{-1}[(m-2)/t_x]$ for $t_y=t_z=1$.}
{Following the same arguments as before, we compute $P(k_{x})/P_0$ across all momenta and find that $n(k_{x}) = 0$ $\forall \abs{k_x}\ge k_0$. This matches the calculated Chern number using the FHS method for those momenta. However, the non-zero values of $n(k_{x})$ for $\abs{k_x}<k_0$ only agree well with the calculated Chern numbers at momenta farther away from the gapless points (cf. blue line labeled $\omega_1$, \cref{fig:CvsK}(b)). This deviation close to the Weyl nodes occurs because, in the proximity of nodes, for finite driving frequency $\omega_{1}$, the system ``unlocks" from the instantaneous state in a short time and undergoes Landau-Zener transitions. As a result of this, the adiabaticity required for quantized energy transfer breaks down. The timescale over which the system remains locked in is inversely proportional to $(\sqrt{\omega_1\omega_2})^{\color{red}3/2}$~\cite{PhysRevLett.125.100601}. Thus, upon decreasing the driving frequency $\omega_1$, this timescale increases, allowing the system to remain in the instantaneous eigenstate for a longer duration. Consequently, the long-time-averaged energy transfer calculated before the system deviates from the instantaneous eigenstate remains quantized, making the step-like behavior become more pronounced as we move closer to the zero-frequency limit, as shown in \cref{fig:CvsK}(b). We next show that the real-space power transfer encodes the projected Weyl-node separation $2k_0$, equal to the corresponding Fermi-arc extent. In order to extract this information, we Fourier transform the Hamiltonian with respect to $k_x$ and impose periodic boundary conditions. For $N$ unit cells, each with two degrees of freedom, the resulting problem is a $2N$-level system on a two-dimensional Floquet lattice}
\begin{align}
\mathcal{H}(t)&=\sum_{i}\psi_{i}^{\dagger} \mu(t)\sigma_{z}\psi_{i}\notag\\
&+\left(\psi_{i}^{\dagger} t_{2}(t)\sigma_{x}\psi_{i}-t_{x}\psi_{i}^{\dagger}\sigma_{x}\psi_{i+1}+\mathrm{h.c.}\right),
\label{eq:7}
\end{align}
where $\psi_i=(\psi_{iA},\psi_{iB})^{T}$ is the annihilation operator for an electron at site $i$. The spinor structure describes the sublattice degree of freedom, with $\mu(t)=2t_z\sin(\omega_2 t)$ and $t_{2}(t)=2\left(m-t_y\cos(\omega_1 t)-t_z\cos(\omega_2 t)\right)-2it_y\sin(\omega_1 t)$. 
{Using the Hamiltonian in \cref{eq:7}, we compute the power transfer between the drives in the \(\mathrm{W}_2\) region from \cref{eq:6}, by setting the half-filled state as the initial state. Drawing parallels from previous works on Hall conductivity in Weyl semimetals \cite{PhysRevLett.107.127205}, we find that $2\pi P/P_{0}$ matches qualitatively with the Fermi arc length as seen in \cref{fig:CvsK}(c).}

{Furthermore, similar to the Weyl phase with two Weyl nodes, we compute the power transfer between the two driving frequencies for Weyl phases with multiple nodes, namely \(4\), \(6\), and \(8\) Weyl nodes. We consider \(m/t_y=0.5\) and vary \(t_x/t_y\) to access these different Weyl phases. We find that, similar to the two-node Weyl phase, the momentum-resolved Chern number extracted from the slope of the power transfer between the two drives agrees well with the Chern number calculated using the FHS method, as shown in~\cref{fig:fig3}. More importantly, unlike the two-node Weyl phase, where the momentum-resolved Chern number changes by \(1\) when crossing a plane containing a Weyl node, we observe a change of \(2\) in the four-node Weyl phase, as shown in~\cref{fig:fig3}(a). This behavior arises because, in the four-node phase, pairs of Weyl nodes with the same chirality lie in the same \(k_y-k_z\) plane. The same reasoning explains the change in the Chern number by \(2\) in the six- and eight-node Weyl phases.
However, in the six- and eight-node phases, there are also \(k_y-k_z\) planes containing only one Weyl node. In such cases, the Chern number changes by \(1\) as one crosses the plane, as shown in~\cref{fig:fig3}(b,c).}

\textit{Experimental relevance.---}
{Recent experiments have established the key ingredients of our proposal: multi-frequency control, synthetic dimensions simulations, topological pumping, and drive-resolved energy measurements.
A single NV-center electronic spin driven by two incommensurate radio-frequency tones exhibited a synthetic quantum Hall response through Chern-number-locked oscillations and Berry-curvature signatures~\cite{PhysRevLett.125.160505}.
Likewise, quasiperiodic driving of an IBM superconducting qubit realized a temporal half-BHZ Chern insulator and its topological transition~\cite{PhysRevLett.126.163602}, building on mature circuit-control capabilities~\cite{krantz2019quantum, RevModPhys.93.025005} and proposals for higher-Chern Floquet lattices~\cite{lei2024simulating}.
In ultracold atoms, experiments have observed Hall-drift reversal at topological boundaries~\cite{science.adg3848}, the four-dimensional quantum Hall response through a two-dimensional pump~\cite{lohse2018exploring}, and Thouless pumping along a radio-frequency-controlled Rydberg synthetic dimension~\cite{trautmann2024realization}.
Photonics provides the most direct access to interdrive energy transfer: an incommensurately driven photonic molecule showed drive-resolved topological pumping signatures and a route to Weyl-point and Berry-curvature measurements~\cite{sridhar2024quantized}.
Single photonic cavities have also realized two independent synthetic dimensions~\cite{dutt2020single}, while a frequency-encoded photonic Chern insulator recently exhibited quantized Hall drift~\cite{chenier2026quantized}.
Together, these advances make the ($k_x$)-resolved pumping plateau and node-separation-dependent total power accessible in qubit, atomic, and photonic settings.

Beyond experimental feasibility, the Weyl-semimetal realization offers a control advantage absent in fully gapped topological frequency converters. In a gapped topological insulator, the adiabatic conversion coefficient is locked to an integer Chern number. Consequently, for fixed drive frequencies, the topological contribution to the power transfer cannot be varied continuously without closing the gap and crossing a topological phase boundary. In the Weyl phase studied here, by contrast, the net power transfer is determined by the separation between the two Weyl nodes. Since this separation can be varied continuously by tuning the system parameters while remaining within the same Weyl phase, the transferred power can likewise be controlled continuously rather than only in discrete topological steps. In the ideal adiabatic limit, this permits arbitrarily fine power-control settings; experimentally, the achievable resolution will be limited by finite-size effects, nonadiabatic corrections, decoherence, and parameter calibration. The tunable Weyl-node separation, therefore, provides a geometric control knob for precision energy transfer between the drives.}

\textit{Outlook.---} {We have shown that in a mixed driven system of one real and two synthetic dimensions, the characteristic topological response of a time-reversal-breaking Weyl semimetal can be realized. Close to the adiabatic regime, the energy transferred between the drives at each fixed conserved momentum ($k_x$) measures the Chern number of the corresponding synthetic two-dimensional slice, such that the boundaries of the quantized pumping plateau locate the Weyl nodes. At finite driving frequencies, Landau-Zener transitions produce deviations from quantization near the gapless points, while the systematic recovery of the plateau upon lowering the frequencies establishes the controlled adiabatic origin of these corrections. For the half-filled real-space system, the net power transfer approaches the projected Weyl-node separation ($2k_0$), providing a bulk measure equal to the Fermi-arc length of the corresponding system. Taken together, the quantized momentum-resolved response and the continuous total response provide complementary dynamical signatures of the same gapless topology. Because the Weyl-node separation can be varied continuously without leaving the Weyl phase, it provides a clean control knob for tuning the energy transferred between the drives beyond the integer-locked topological conversion of gapped phases. Recent experiments using driven cold atoms~\cite{science.adg3848}, quasiperiodically driven qubits, and photonic synthetic dimensions~\cite{PhysRevLett.125.160505, sridhar2024quantized} suggest that both responses are experimentally accessible, positioning mixed Floquet lattices as a platform for realizing, diagnosing, and controlling gapless topological matter.}

\textit{Acknowledgments---} The authors thank Anushya Chandran, Philip JD Crowley, Bertrand Halperin, Chris Laumann, and Suhas Gangadharaiah for insightful discussions and input. Ashutosh acknowledges support from the ANRF (Government of India) via Sanction No.~PDF/2025/001247, IISER Tirupati. Ankur thanks IISER Tirupati for support, ANRF for start-up grant ANRF/ECRG/2024/001172/PMS, and DST for grant DST/NPNST/AM/2025/31.
\bibliographystyle{apsrev4-1}
\bibliography{refs}
\end{document}